\documentstyle[amssymb,floats,aps,prb,epsf]{revtex}
\epsfclipon
\begin{document}
\twocolumn[\hsize\textwidth\columnwidth\hsize\csname
@twocolumnfalse\endcsname

\draft

\title{Kinetic energy driven superconductivity in the electron
doped cobaltate Na$_{x}$CoO$_{2}\cdot y$H$_{2}$O}
\author{Bin Liu, Ying Liang, and Shiping Feng}
\address{Department of Physics, Beijing Normal University,
Beijing 100875, China}

\author{Wei Yeu Chen}
\address{Department of Physics, Tamkang University, Tamsui 25137,
Taiwan}

\maketitle
\begin{abstract}
Within the charge-spin separation fermion-spin theory, we have
shown that the mechanism of superconductivity in the electron
doped cobaltate Na$_{x}$CoO$_{2}\cdot y$H$_{2}$O is ascribed to
its kinetic energy. The dressed fermions interact occurring
directly through the kinetic energy by exchanging magnetic
excitations. This interaction leads to a net attractive force
between dressed fermions, then the electron Cooper pairs
originating from the dressed fermion pairing state are due to the
charge-spin recombination, and their condensation reveals the
superconducting ground state. The superconducting transition
temperature is identical to the dressed fermion pair transition
temperature, and is suppressed to a lower temperature due to the
strong magnetic frustration. The optimal superconducting
transition temperature occurs in the electron doping concentration
$\delta\approx 0.29$, and then decreases for both underdoped and
overdoped regimes, in qualitative agreement with the experimental
results.
\end{abstract}
\

 ]
\bigskip

\narrowtext

The discovery of superconductivity in the doped cobaltate
Na$_{0.35}$CoO$_{2}\cdot 1.3$H$_{2}$O has engendered great
interest in transition metal oxides \cite{takada}. Although the
superconducting (SC) transition temperature is much less than that
in the doped cuprate superconductors, both systems share many
common features. The doped cobaltate has a lamellar structure
consisting of the two-dimensional (2D) CoO$_{2}$ layers separated
by a thick insulating layer of Na$^{+}$ ions, where the one-half
spin Co$^{4+}$ ions are arranged on a {\it triangular lattice},
and it becomes a superconductor after hydration that significantly
enhances the distance between CoO$_{2}$ layers \cite{takada}. This
structure is similar to the cuprate superconductors in the sense
that they also have a layered structure of the {\it square
lattice} of the CuO$_{2}$ plane separated by insulating layers
\cite{kastner}. The undoped cuprate is a Mott insulator with the
antiferromagnetic (AF) long-range-order (AFLRO), then changing the
carrier concentration by ionic substitution or increasing the
oxygen content turns these compounds into the SC state leaving the
AF short-range correlation still intact \cite{kastner,tokura}. The
optimal doping occurs at $0.15$ holes (electrons) added to the
half-filled band of the parent compound \cite{kastner,tokura}.
However, Na$_{x}$CoO$_{2}\cdot y$H$_{2}$O is viewed as an electron
doped Mott insulator, where superconductivity appears with
electron doping \cite{takada}. The optimal doping for
superconductivity occurs at $0.3$ electrons per Co above the
ground-state Na$_{0.3}$CoO$_{2}\cdot 1.3$H$_{2}$O, which is a
half-filled two-electron $t_{2g}$ derived band
\cite{schaak,uemura2}. Although the ferromagnetic correlation is
present in Na$_{x}$CoO$_{2}$ for the large electron doping
concentration ($x\approx 0.7$) \cite{motohashi}, the AF
short-range spin correlation in Na$_{x}$CoO$_{2}$ and
Na$_{x}$CoO$_{2}\cdot y$H$_{2}$O in the low electron doping
concentration ($\delta\approx 0.35$) has been observed from
nuclear quadrupolar resonance and thermopower as well as other
experimental measurements \cite{takada,schaak,uemura2,chu,wang}.
Moreover, the SC transition temperature displays the same kind of
doping controlled behavior that is observed in the doped cuprates,
i.e., the optimal SC transition temperature occurs in a narrow
range of the electron doping concentration, and decreases for both
underdoped and overdoped regimes \cite{schaak,uemura2}, in analogy
to the phase diagram of the cuprate superconductors
\cite{uemura,tallon}. Therefore the superconductivity developed
with the background of the AF spin correlation in the doped
cobaltate suggests the importance of the strong electron
correlation as in the doped cuprates
\cite{kastner,tokura,schaak,chu}.

In the conventional superconductors, the electrons interact by
exchanging phonons. Since this interaction leads to a net
attractive force between electrons, then the system can lower its
potential energy by forming electron Cooper pairs
\cite{schrieffer}. These electron Cooper pairs condense into a
coherent macroscopic quantum state, then can move freely without
resistance. As a result, pairing in the conventional
superconductors is always related with an increase in kinetic
energy which is overcompensated by the lowering of potential
energy \cite{schrieffer}. On the contrary, it has been argued that
the SC transition in the doped cuprates is determined by the need
to reduce the frustrated kinetic energy
\cite{anderson1,molegraaf}, where the driving attractive force
between holes may be mediated by the exchange of spin excitations
\cite{dagotto}. Since there is a remarkable resemblance in the
normal- and SC-state properties between Na$_{x}$CoO$_{2}\cdot
y$H$_{2}$O and hole doped cuprates
\cite{takada,kastner,tokura,schaak,chu} as mentioned above, and
the strong electron correlation is common for both these
materials, then two systems may have similar underlying SC
mechanism, i.e., it is possible that superconductivity in the
electron doped cobaltates is also driven by the kinetic energy as
in the doped cuprates. Within the charge-spin separation
fermion-spin theory \cite{feng1}, the mechanism of
superconductivity in the hole doped cuprates has been discussed
\cite{feng2}, where dressed holons interact occurring directly
through the kinetic energy by exchanging the dressed spinon
excitations, leading to a net attractive force between the dressed
holons, then the electron Cooper pairs originating from the
dressed holon pairing state are due to the charge-spin
recombination, and their condensation reveals the SC ground-state.
In this case, the electron SC transition temperature is determined
by the dressed holon pair transition temperature, and is a
nonmonotonic doping dependence, i.e., the SC transition
temperature is maximal for a particular carrier concentration and
decreases both for underdoped and overdoped regimes \cite{feng2},
in agreement with experiments \cite{uemura,tallon}. In this paper,
we study the mechanism of superconductivity in the doped
cobaltates along with this line. We find that the SC transition
temperature in the electron doped cobaltates is suppressed to a
lower temperature due to the strong magnetic frustration.

In the electron doped cobaltates, the characteristic feature is
the presence of the 2D CoO$_{2}$ plane \cite{takada}, then the
unusual behaviors are dominated by this plane. It has been argued
that the essential physics of the doped CoO$_{2}$ plane is
contained in the $t$-$J$ model on a triangular lattice
\cite{baskaran},
\begin{eqnarray}
H&=&t\sum_{i\hat{\eta}\sigma}PC_{i\sigma}^{\dagger}
C_{i+\hat{\eta}\sigma}P^{\dagger}-\mu\sum_{i\sigma}P
C_{i\sigma }^{\dagger}C_{i\sigma }P^{\dagger} \nonumber \\
&+&J\sum_{i\hat{\eta}}{\bf S}_{i}\cdot{\bf S}_{i+\hat{\eta}},
\end{eqnarray}
where $t<0$, the summation is over all sites $i$, and for each
$i$, over its nearest-neighbor $\hat{\eta}$,
$C^{\dagger}_{i\sigma}$ ($C_{i\sigma}$) is the electron creation
(annihilation) operator, ${\bf S}_{i}=C^{\dagger}_{i}{\bf
\sigma}C_{i}/2$ is the spin operator with ${\bf
\sigma}=(\sigma_{x},\sigma_{y},\sigma_{z})$ as the Pauli matrices,
$\mu$ is the chemical potential, and the projection operator $P$
removes zero occupancy, i.e., $\sum_{\sigma}C^{\dagger}_{i\sigma}
C_{i\sigma}\geq 1$. The strong electron correlation manifests
itself by the local constraint \cite{anderson2}. For a proper
treatment of the single occupancy local constraint in the doped
cuprates in analytical calculations, the charge-spin separation
fermion-spin theory has been developed to incorporate the
constraint \cite{feng1}. To apply this theory in the electron
doped cobaltates, the $t$-$J$ model (1) can be rewritten in terms
of a particle-hole transformation $C_{i\sigma}\rightarrow
f^{\dagger}_{i-\sigma}$ as,
\begin{eqnarray}
H=-t\sum_{i\hat{\eta}\sigma}f_{i\sigma}^{\dagger}
f_{i+\hat{\eta}\sigma}+\mu\sum_{i\sigma}f_{i\sigma }^{\dagger}
f_{i\sigma }+J\sum_{i\hat{\eta}}{\bf S}_{i} \cdot{\bf
S}_{i+\hat{\eta}},
\end{eqnarray}
supplemented by the local constraint
$\sum_{\sigma}f^{\dagger}_{i\sigma}f_{i\sigma}\leq 1$ to remove
double occupancy, where $f^{\dagger}_{i\sigma}$ ($f_{i\sigma}$) is
the hole creation (annihilation) operator, and ${\bf
S}_{i}=f^{\dagger}_{i}{\bf \sigma}f_{i}/2$ is the spin operator in
the hole representation. Then the hole operators can be expressed
as, $f_{i\uparrow}=a^{\dagger}_{i\uparrow}S^{-}_{i}$ and
$f_{i\downarrow}=a^{\dagger}_{i\downarrow}S^{+}_{i}$, in the
charge-spin separation fermion-spin theory \cite{feng1}, where the
spinful fermion operator $a_{i\sigma}=e^{-i\Phi_{i\sigma}}a_{i}$
describes the charge degree of freedom together with some effects
of the spin configuration rearrangements due to the presence of
the doped electron itself (dressed fermion), while the spin
operator $S_{i}$ describes the spin degree of freedom (dressed
spinon), then the single occupancy local constraint,
$\sum_{\sigma} f^{\dagger}_{i\sigma}f_{i\sigma}
=S^{+}_{i}a_{i\uparrow} a^{\dagger}_{i\uparrow} S^{-}_{i}+
S^{-}_{i}a_{i\downarrow} a^{\dagger}_{i\downarrow} S^{+}_{i}=
a_{i}a^{\dagger}_{i} (S^{+}_{i} S^{-}_{i}+S^{-}_{i} S^{+}_{i})=1-
a^{\dagger}_{i} a_{i}\leq 1$, is satisfied in analytical
calculations, and the double dressed fermion occupancy,
$a^{\dagger}_{i\sigma} a^{\dagger}_{i-\sigma}= e^{i\Phi_{i\sigma}}
a^{\dagger}_{i} a^{\dagger}_{i} e^{i\Phi_{i-\sigma}}=0$ and
$a_{i\sigma} a_{i-\sigma}= e^{-i\Phi_{i\sigma}}a_{i}a_{i}
e^{-i\Phi_{i-\sigma}} =0$, are ruled out automatically. These
dressed fermion and spinon have been shown to be gauge invariant,
and in this sense, they are real and can be interpreted as the
physical excitations \cite{feng1,laughlin}. It has been emphasized
that this dressed fermion $a_{i\sigma}$ is a spinless fermion
$a_{i}$ incorporated a spin cloud $e^{-i\Phi_{i\sigma}}$ (magnetic
flux), and is a magnetic dressing. In other words, the gauge
invariant dressed fermion carries some spin messages, i.e., it
shares its nontrivial spinon environment \cite{martins}. Although
in common sense $a_{i\sigma}$ is not a real spinful fermion, it
behaves like a spinful fermion. In this charge-spin separation
fermion-spin representation, the low-energy behavior of the
$t$-$J$ model (2) can be expressed as \cite{feng1},
\begin{eqnarray}
H&=&-t\sum_{i\hat{\eta}}(a_{i\uparrow}S^{+}_{i}
a^{\dagger}_{i+\hat{\eta}\uparrow}S^{-}_{i+\hat{\eta}}+
a_{i\downarrow}S^{-}_{i}a^{\dagger}_{i+\hat{\eta}\downarrow}
S^{+}_{i+\hat{\eta}}) \nonumber \\
&-&\mu\sum_{i\sigma}a^{\dagger}_{i\sigma} a_{i\sigma}+J_{{\rm
eff}}\sum_{i\hat{\eta}}{\bf S}_{i}\cdot {\bf S}_{i+\hat{\eta}},
\end{eqnarray}
with $J_{{\rm eff}}=(1-\delta)^{2}J$, and $\delta=\langle
a^{\dagger}_{i\sigma}a_{i\sigma}\rangle=\langle a^{\dagger}_{i}
a_{i}\rangle$ is the electron doping concentration. In this case,
the magnetic energy ($J$) term in the $t$-$J$ model is only to
form an adequate dressed spinon configuration \cite{anderson1},
while the kinetic energy ($t$) term has been transferred as the
dressed fermion-spinon interaction, which dominates the essential
physics. This dressed fermion-spinon interaction is quite strong,
and can induce the dressed fermion pairing state (then the
electron pairing state and superconductivity) by exchanging
dressed spinon excitations in a higher power of the electron
doping concentration $\delta$ as in the doped cupares
\cite{feng2}. As in the conventional superconductors, the SC state
in the doped cobaltates also is characterized by electron Cooper
pairs, forming SC quasiparticles, then the order parameter for the
electron Cooper pair can be expressed as,
\begin{eqnarray}
\Delta=\langle f^{\dagger}_{i\uparrow}f^{\dagger}_{j\downarrow}-
f^{\dagger}_{i\downarrow}f^{\dagger}_{j\uparrow}\rangle =\langle
a_{i\uparrow}a_{j\downarrow}S^{+}_{i}S^{-}_{j}-a_{i\downarrow}
a_{j\uparrow}S^{-}_{i}S^{+}_{j}\rangle .
\end{eqnarray}
At the zero doping, the $t$-$J$ model is reduced to the Heisenberg
model. Many authors \cite{caprioti} have shown unambiguously that
as in a square lattice, there is indeed AFLRO in the ground state
of the AF Heisenberg model on a triangular lattice. However, this
AFLRO is destroyed more rapidly with increasing doping than that
on a square lattice due to the strong geometry frustration. Thus
there is no AFLRO away from the zero doping, i.e., $\langle
S_{i}^{z}\rangle=0$, where the dressed spinons form the disordered
spin liquid state, then the dressed spinon correlation function
$\langle S^{+}_{i}S^{-}_{j}\rangle=\langle S^{-}_{i} S^{+}_{j}
\rangle$, and the order parameter for the electron Cooper pair in
Eq. (4) can be written as $\Delta=-\langle S^{+}_{i}S^{-}_{j}
\rangle\Delta_{a}$, with the dressed fermion pairing order
parameter $\Delta_{a}=\langle a_{j\downarrow} a_{i\uparrow}-
a_{j\uparrow}a_{i\downarrow}\rangle$. This shows that the dressed
fermion pairs move freely in the background of the disordered spin
liquid state, and then the physical properties of the SC state are
essentially determined by the dressed fermion pairing state. In
this case, the mechanism of superconductivity in the doped
cuprates has been discussed \cite{feng2} based on the Eliashberg's
strong coupling theory \cite{eliashberg}. Following the
discussions in Ref. \cite{feng2}, we obtain the self-consistent
equations in terms of the equation of motion method that is
satisfied by the full dressed fermion diagonal and off-diagonal
Green's functions $g(i-j,t-t') =\langle\langle
a_{i\sigma}(t);a^{\dagger}_{j\sigma}(t')\rangle \rangle$ and
$\Im^{\dagger} (i-j,t-t')=\langle\langle
a^{\dagger}_{i\uparrow}(t); a^{\dagger}_{j\downarrow}(t')
\rangle\rangle$ as,
\begin{mathletters}
\begin{eqnarray}
g(i\omega_{n},k)&=&g^{(0)}(i\omega_{n},k)+g^{(0)}(i\omega_{n},k)
[\Sigma^{(a)}_{1}(i\omega_{n},k) \nonumber\\
&\times& g(i\omega_{n},k) -\Sigma^{(a)}_{2}
(-i\omega_{n},-k)\Im^{\dagger}(i\omega_{n},k)], \\
\Im^{\dagger}(i\omega_{n},k)&=&g^{(0)}(-i\omega_{n},-k)
[\Sigma^{(a)}_{1}(-i\omega_{n},-k)
\Im^{\dagger}(-i\omega_{n},-k)\nonumber\\
&+&\Sigma^{*(a)}_{2} (-i\omega_{n},-k)g(i\omega_{n},k)],
\end{eqnarray}
\end{mathletters}
respectively, where the mean-field (MF) dressed fermion Green's
function $g^{(0)-1}(i\omega_{n},k) =i\omega_{n}-\xi_{k}$, with the
MF dressed fermion excitation spectrum $\xi_{k}=Zt\chi\gamma_{k}
-\mu$, $Z$ is the number of the nearest neighbor sites,
$\gamma_{k}=(1/Z)\sum_{\hat{\eta}}e^{i{\bf k}\cdot\hat{\eta}}=
[{\rm cos}k_{x}+2{\rm cos}(k_{x}/2){\rm cos}(\sqrt{3}k_{y}/2)]
/3$, the dressed spinon correlation function $\chi=\langle
S_{i}^{+} S_{i+\hat{\eta}}^{-} \rangle$, while the self-energies
have been obtained as,
\begin{mathletters}
\begin{eqnarray}
\Sigma^{(a)}_{1}(i\omega_{n},k)&=&(Zt)^{2}{1\over N^{2}} \nonumber\\
&\times& \sum_{{\bf p,p'}}\gamma^{2}_{{\bf p+p'+k}}{1\over
\beta}\sum_{ip_{m}}
g(ip_{m}+i\omega_{n},p+k)\nonumber \\
&\times& {1\over\beta}\sum_{ip'_{m}}D^{(0)}(ip'_{m},p')\nonumber\\
&\times& D^{(0)}(ip'_{m}+ip_{m},p'+p), \\
\Sigma^{(a)}_{2}(i\omega_{n},k)&=&(Zt)^{2}{1\over N^{2}} \nonumber\\
&\times& \sum_{{\bf p,p'}}\gamma^{2}_{{\bf p+p'+k}}{1\over \beta}
\sum_{ip_{m}}\Im
(-ip'_{m}-i\omega_{n},-p-k)\nonumber\\
&\times& {1\over\beta}\sum_{ip'_{m}}D^{(0)}(ip'_{m},p') \nonumber\\
&\times& D^{(0)}(ip'_{m}+ip_{m},p'+p),
\end{eqnarray}
\end{mathletters}
where the MF dressed spinon Green's function $D^{(0)}(i-j,t-t')
=\langle\langle S^{+}_{i}(t);S^{-}_{j}(t')\rangle\rangle_{0}$, and
has been obtained as $D^{(0)-1}(ip_{m},p)=[(ip_{m})^{2}-
\omega_{p}^{2}]/B_{p}$, with $B_{p}=\lambda [2 \chi^{z}(\epsilon
\gamma_{p}-1)+\chi (\gamma_{p}-\epsilon)]$, the MF dressed fermion
excitation spectrum $\omega^{2}_{ p}=A_{1} \gamma^{2}_{p}+A_{2}
\gamma_{p}+A_{3}$, $A_{1}=\alpha\epsilon \lambda^{2}(\epsilon
\chi^{z}+\chi/2)$, $A_{2}=-\epsilon \lambda^{2}[\alpha(\chi^{z}+
\epsilon\chi/2)+(\alpha C^{z}+ (1-\alpha)/(4Z)-\alpha\epsilon
\chi/(2Z))+(\alpha C+(1-\alpha) /(2Z)-\alpha\chi^{z}/2)/2]$,
$A_{3}=\lambda^{2}[\alpha C^{z}+ (1-\alpha)/(4Z)-\alpha\epsilon
\chi/(2Z)+\epsilon^{2}(\alpha C+ (1-\alpha)/(2Z)-\alpha
\chi^{z}/2)/2]$, $\lambda=2ZJ_{eff}$, $\epsilon=1+2t\phi/J_{{\rm
eff}}$, the dressed fermion's particle-hole parameters $\phi=
\langle a^{\dagger}_{i\sigma} a_{i+\hat{\eta}\sigma}\rangle$, and
the dressed spinon correlation functions $\chi^{z}=\langle
S_{i}^{z}S_{i+\hat{\eta}}^{z}\rangle$, $C=(1/Z^{2})
\sum_{\hat{\eta},\hat{\eta'}}\langle S_{i+\hat{\eta}}^{+}
S_{i+\hat{\eta'}}^{-}\rangle$, and $C^{z}=(1/Z^{2})
\sum_{\hat{\eta},\hat{\eta'}}\langle S_{i+\hat{\eta}}^{z}
S_{i+\hat{\eta'}}^{z}\rangle$. In order not to violate the sum
rule of the correlation function $\langle S^{+}_{i}S^{-}_{i}
\rangle=1/2$ in the case without AFLRO, the important decoupling
parameter $\alpha$ has been introduced in the MF calculation
\cite{kondo,feng3}, which can be regarded as the vertex
correction.

Since the pairing force and dressed fermion gap function have been
incorporated into the self-energy $\Sigma^{(a)}_{2}(k)$, then it
is an effective dressed fermion gap function. In particular, this
effective dressed fermion gap function $\Sigma^{(a)}_{2}(k)$ is an
even function of $i\omega_{n}$, while the other self-energy
$\Sigma^{(a)}_{1}(k)$ is not. It has been shown \cite{liu} that
the self-energy $\Sigma^{(a)}_{1}(k)$ renormalizes the MF dressed
fermion spectrum, and therefore it dominates the charge transport
of the systems. As a qualitative discussion as in the doped
cuprates \cite{feng2}, we neglect $\Sigma^{(a)}_{1}(k)$, and only
study the static limit of the effective dressed fermion gap
function, i.e., $\Sigma^{(a)}_{2}(k)=\bar{\Delta}_{a}({\bf k})$,
then the dressed fermion diagonal and off-diagonal Green's
functions are obtained from Eq. (5) as,
\begin{mathletters}
\begin{eqnarray}
g(i\omega_{n},k)&=&{i\omega_{n}+\xi_{k}\over (i\omega_{n})^{2}-
E^{2}_{k}}\nonumber \\
&=&{1\over 2}\sum_{\nu=1,2}\left ( 1+{\xi_{k}\over
E_{\nu}(k)} \right ){1\over i\omega_{n}-E_{\nu}(k)},\\
\Im^{\dagger}(i\omega_{n},k)&=&-{\bar{\Delta}^{*}_{a}(k)\over
(i\omega_{n})^{2}-E^{2}_{k}}\nonumber \\
&=&-{1\over 2}\sum_{\nu=1,2} {\bar{\Delta}^{*}_{a}(k)\over
E_{\nu}(k)}{1\over i\omega_{n} -E_{\nu}(k)},
\end{eqnarray}
\end{mathletters}
with $E_{1}(k)=E_{k}$, $E_{2}(k)=-E_{k}$, and the dressed fermion
quasiparticle spectrum $E_{k}= \sqrt{\xi^{2}_{ k}+ \mid
\bar{\Delta}_{a}(k)\mid^{2}}$. Although the situation of the
pairing symmetry of superconductivity in the doped cobaltate
Na$_{x}$CoO$_{2}\cdot y$H$_{2}$O is far from reaching a consensus,
many experimental data suggest non-s-wave superconductivity
without a full gap \cite{takada,schaak,chu,wang}. In particular,
it has been argued according to the irreducible representations of
the triangular lattice that there are three possible basis
functions of even parity \cite{feng4}, i.e., one s-like function
$s_{k}={\rm cos} k_{x}+{\rm cos}[(k_{x}- \sqrt{3} k_{y})/2]+{\rm
cos} [(k_{x}+ \sqrt{3}k_{y})/2]$, and two d-like functions,
$d_{1k}= 2{\rm cos}k_{x}-{\rm cos}[(k_{x}- \sqrt{3} k_{y})/2]-{\rm
cos}[(k_{x}+ \sqrt{3}k_{y})/2]$ and $d_{2k}=\sqrt{3}{\rm cos}
[(k_{x}+\sqrt{3}k_{y})/2] -\sqrt{3}{\rm cos} [(k_{x}-
\sqrt{3}k_{y})/2]$. However, with the different linear
combinations of these basis functions, it has been found
\cite{feng4} in terms of the Gutzwiller approximation scheme and
variational Monte Carlo simulation that the lowest energy state is
the d-wave $(d_{1}+ id_{2})$ state with the gap function
$\Delta^{(d)}(k)\propto \Delta^{(d)}(d_{1k}+id_{2k})$. Recently,
many authors have argued that it is possible that this d-wave
state also is the lowest state around the electron doped regime
where superconductivity appears \cite{baskaran2}. Therefore in the
following discussions, we only consider this d-wave case
$\bar{\Delta}^{(d)}_{a}(k)= \bar{\Delta}^{(d)}_{a} (d_{1k}+
id_{2k})$. In this case, we find from Eq. (6b) that the effective
dressed fermion gap parameter satisfies the equation,
\begin{eqnarray}
1&=&-(Zt)^{2}{1\over N^{3}}\sum_{k,q,p}\gamma^{2}_{k+q}
\gamma^{(d)}_{k-p+q}\gamma^{(d)}_{k}\sum_{\nu,\nu',\nu''} {1\over
2E_{\nu''}(k)}\nonumber \\
&\times& {B_{q}B_{p}\over\omega_{\nu}(q)\omega_{\nu'}(p)}
{F_{\nu\nu'\nu''}(k,q,p)\over \omega_{\nu'}(p)-\omega_{\nu}(q)
-E_{\nu''}(k)},
\end{eqnarray}
where $\gamma^{(d)}_{k}=d_{1k}+id_{2k}$, $F_{\nu\nu'\nu''}(k,q,p)
=n_{F}[E_{\nu''}(k)](n_{B} [\omega_{\nu}(q)]-n_{B}[
\omega_{\nu'}(p)])+ n_{B}[\omega_{\nu'}( p)](1+n_{B}[
\omega_{\nu}(q)])$, with $n_{B}[\omega_{\nu}(p)]$ and $n_{F}
[E_{\nu}(k)]$ are the boson and fermion distribution functions,
respectively. As in the case of the doped cuprates \cite{feng2},
this gap equation must be solved simultaneously with other seven
self-consistent equations, therefore all the order parameters,
decoupling parameter $\alpha$, and chemical potential $\mu$ are
determined by the self-consistent calculation \cite{feng3}. In
this case, we can obtain the dressed fermion pair order parameter
in terms of the off-diagonal Green's function (7b) as,
\begin{eqnarray}
\Delta^{(d)}_{a}={2\over N}\sum_{k}\mid\gamma^{(d)}_{k}\mid^{2}
{\bar{\Delta}^{(d)}_{a}\over E_{k}}{\rm th} [{1\over 2}\beta E_{
k}].
\end{eqnarray}
The dressed fermion pairing state originating from the kinetic
energy term by exchanging dressed spinon excitations will also
lead to form the electron Cooper pairing state as mentioned in Eq.
(4). For a discussion of the physical properties of the SC state,
we now need to calculate the electron off-diagonal Green's
function $\Gamma^{\dagger}(i-j,t-t')=\langle\langle
f^{\dagger}_{i\uparrow}(t);f^{\dagger}_{j\downarrow}(t')\rangle
\rangle$. In the framework of the charge-spin separation
fermion-spin theory, it is a convolution of the dressed spinon
Green's function $D(ip_{m},p)$ and off-diagonal dressed fermion
Green's function $\Im(i\omega_{n},k)$, and can be expressed as,
\begin{eqnarray}
\Gamma^{\dagger}(i\omega_{n},k)&=&{1\over N}\sum_{p}{1\over \beta}
\sum_{ip_{m}}D^{(0)}(ip_{m},p) \nonumber \\
&\times& \Im(ip_{m}-i\omega_{n},p-k),
\end{eqnarray}
in the present approximation. This convolution of the dressed
spinon Green's function and off-diagonal dressed fermion Green's
function reflects the charge-spin recombination \cite{anderson1},
and can be evaluated in terms of the MF dressed spinon Green's
function and off-diagonal dressed fermion Green's function as,
\begin{eqnarray}
\Gamma^{\dagger}(k)&=&-{1\over N}\sum_{p,\nu,\nu'}
{\bar{\Delta}^{(d)}_{a}(p-k)\over 2E_{\nu'}(p-k)}
\nonumber \\
&\times& {B_{p}\over 2\omega_{\nu}(p)}{n_{B}
[\omega_{\nu}(p)]+n_{F}[E_{\nu'}(p-k)]\over
i\omega_{n}-\omega_{\nu}(p)+E_{\nu'}(p-k)}.
\end{eqnarray}
In this case, the SC gap function is obtained in terms of this
electron off-diagonal Green's function as,
\begin{eqnarray}
\Delta^{(d)}(k)&=&-{1\over N}\sum_{p}
{\bar{\Delta}^{(d)}_{a}(p-k)\over 2E_{p-k}}{\rm th} [{1\over
2}\beta E_{p-k}]{B_{p}\over 2\omega_{p}}
\nonumber \\
&\times& {\rm coth}[{1\over 2}\beta\omega_{p}],
\end{eqnarray}
\begin{figure}[prb]
\epsfxsize=3.0in\centerline{\epsffile{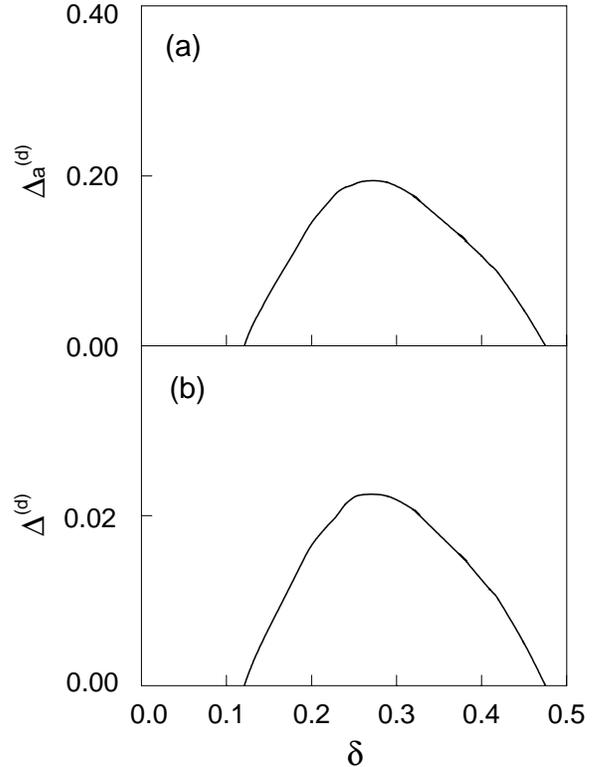}} \caption{The
dressed fermion (a) and superconducting (b) gap parameters in the
d-wave symmetry as a function of the electron doping concentration
in temperature $T=0.001J$ and $t/J=-2.5$.}
\end{figure}
which shows that the symmetry of the electron Cooper pair is the
same as the symmetry of the dressed fermion pair, i.e., the SC gap
function can be written as $\Delta^{(d)}(k)=\Delta^{(d)}
(d_{1k}+id_{2k})$. With the help of Eqs. (12) and (9), the SC gap
parameter can be obtained as $\Delta^{(d)}=-\chi\Delta^{(d)}_{a}$.
As in the doped cuprates \cite{feng2}, our present theory also
indicates that there is a coexistence of the electron Cooper pair
and short-range AF correlation in the doped cobaltates, and
therefore the short-range AF fluctuation can persist into
superconductivity. This is because that the AF fluctuation is
dominated by the scattering of dressed spinons \cite{feng1}, which
has been incorporated into the electron off-diagonal Green's
function (and hence the electron Cooper pair) in terms of the
dressed spinon Green's function. This result is consistent with
some experiments \cite{takada,schaak,uemura2,chu,wang}. We have
perform a numerical calculation for the gap parameters, and the
results of the dressed fermion (a) and SC (b) gap parameters in
the d-wave symmetry as a function of the electron doping
concentration $\delta$ at $T=0.001J$ and $t/J=-2.5$ are shown in
Fig. 1, where both values of the dressed fermion and SC gap
parameters increase with increasing doping in the underdoped
regime, and reach the maximal vaules for a particular doping
concentration, then decrease in the overdoped regime. Although
there is a coexistence of the electron Cooper pair and short-range
AF correlation, the value of the SC gap parameter is still
suppressed by this AF fluctuation.
\begin{figure}[prb]
\epsfxsize=3.0in\centerline{\epsffile{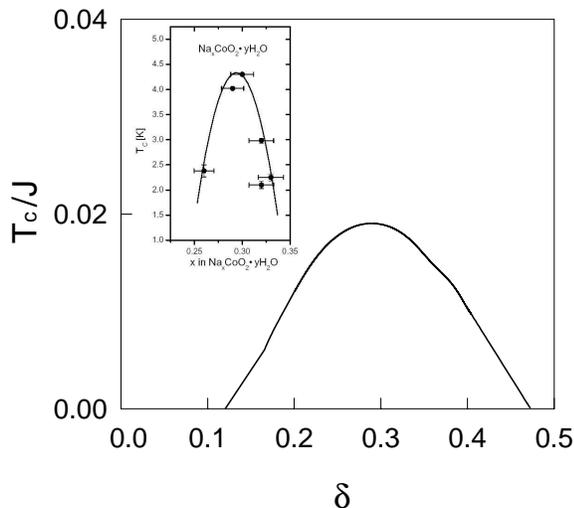}} \caption{The
superconducting transition temperature as a function of the
electron doping concentration in the d-wave symmetry for
$t/J=-2.5$. Inset: the experimental result on
Na$_{x}$CoO$_{2}\cdot y$H$_{2}$O taken from Ref. 4}
\end{figure}
The SC gap function in Eq. (12) also indicates that the SC
transition temperature $T_{c}$ occurring in the case of
$\Delta^{(d)}=0$ is identical to the dressed fermion pair
transition temperature occurring in the case of
$\bar{\Delta}^{(d)}_{a}=0$. This SC transition temperature $T_{c}$
as a function of electron doping concentration $\delta$ in the
d-wave symmetry for $t/J=-2.5$ is plotted in Fig. 2 in comparison
with the experimental data \cite{schaak} taken from
Na$_{x}$CoO$_{2}\cdot y$H$_{2}$O (inset). It is shown that the
maximal SC transition temperature T$_{c}$ occurs around the
optimal electron doping concentration $\delta\approx 0.29$, and
then decreases for both underdoped and overdoped regimes. Using an
reasonable estimation value of $J\sim 10$mev to 20mev in
Na$_{x}$CoO$_{2}\cdot y$H$_{2}$O, the SC transition temperature in
the optimal doping is T$^{{\rm optimal}}_{c}\approx 0.02J\approx
3{\rm K}\sim 6{\rm K}$. Our these results are in qualitative
agreement with the experimental data \cite{schaak,uemura2}.

Since Na$_{x}$CoO$_{2}\cdot y$H$_{2}$O is the electron doped Mott
insulator on a triangular lattice, therefore the system has strong
geometrical spin frustration. This magnetic frustration also
induces the strong charged carrier's quantum fluctuation. In
comparison with the case in the doped cuprates, we find that the
SC transition temperature in the doped cobaltates is suppressed
heavily to a lower temperature due to both strong magnetic
frustration and dressed fermion's quantum fluctuation. In Ref.
\cite{feng2}, the dressed holon's quantum fluctuation is not
considered, and therefore leads to an obvious weakness that the SC
transition temperature in the doped cuprates is too high, and not
suppressed in the overdoped regime. Recently, this weakness has
been cured by considering the dressed holon's quantum fluctuation
\cite{feng5}, and the results show that the SC transition
temperature in the doped cuprates in the optimal doping is
suppressed to an reasonable value, and then decreases for both
underdoped and overdoped regimes, which also confirm that both
strong magnetic frustration and charged carrier's quantum
fluctuation suppress the SC transition temperature.

In summary, we have discussed the mechanism of superconductivity
in the electron doped cobaltate Na$_{x}$CoO$_{2}\cdot y$H$_{2}$O
within the charge-spin separation fermion-spin theory. It is shown
that the mechanism of superconductivity is ascribed to the kinetic
energy. The dressed fermions interact occurring directly through
the kinetic energy by exchanging magnetic excitations. This
interaction leads to a net attractive force between dressed
fermions, then the electron Cooper pairs originating from the
dressed fermion pairing state are due to the charge-spin
recombination, and their condensation reveals the SC ground state.
The electron SC transition temperature is identical to the dressed
fermion pair transition temperature, and is suppressed to a lower
temperature due to the strong magnetic frustration. The optimal SC
transition temperature occurs in the electron doping concentration
$\delta\approx 0.29$, and then decreases for both underdoped and
overdoped regimes, in qualitative agreement with the experimental
observations \cite{schaak,uemura2}.

\acknowledgments This work was supported by the National Natural
Science Foundation of China under Grant Nos. 10125415, 90403005,
10347102, and 10404001, the Grant from Beijing Normal University,
and the National Science Council.

\end{document}